\documentclass{IOS-Book-Article}

\usepackage{mathptmx}
\usepackage{soul}\setuldepth{article}
\def\hb{\hbox to 11.5 cm{}}

\begin{document}

\pagestyle{headings}
\def\thepage{}
\begin{frontmatter}              

\title{A research infrastructure for generating and sharing diversity-aware data}

\markboth{}{}

\author[A]{\fnms{Matteo} \snm{Busso}\orcid{0000-0002-3788-0203}%
\thanks{Corresponding Author: Matteo Busso, PhD Candidate,
email: \url{matteo.busso@unitn.it}}},
\author[B]{\fnms{Ronald} \snm{Chenu Abente Acosta}\orcid{0000-0002-1121-0287}},
\author[B]{\fnms{Amalia} \snm{de Götzen}\orcid{0000-0001-7214-5856}}

\runningauthor{Matteo Busso et al.}
\address[A]{Department of Information Engineering and Computer Science, University of Trento}
\address[B]{Department of Architecture, Design and Media Technology, Aalborg University}

\begin{abstract}
The intensive flow of personal data associated with the trend of computerizing aspects of people's diversity in their daily lives is associated with issues concerning not only people protection and their trust in new technologies, but also bias in the analysis of data and problems in their management and reuse. Faced with a complex problem, the strategies adopted, including technologies and services, often focus on individual aspects, which are difficult to integrate into a broader framework, which can be of effective support for researchers and developers. Therefore, we argue for the development of an end-to-end research infrastructure (RI) that enables trustworthy diversity-aware data within a citizen science community.
\end{abstract}

\end{frontmatter}
\markboth{}{}

\section{Introduction: the relevance of a diversity aware-RI}
Digitization \cite{kitchin2014data} is exponentially increasing the production of data and driving relevant economic, social, and political changes \cite{loebbecke2015reflections}. The trend is often associated with datafication \cite{mayer2013big}, namely the act of quantifying and computerising people everyday life and that, according to \cite{berry2011computational}, has the potential to shape both the ontologies and the methodologies of many disciplines.

However, as \cite{boyd2012critical} would say, "bigger data are not always better data". Indeed, many useful data to model people everyday life as often missing \cite{fielding2008sage}, which leads to the problem of reducing people to their average, ignoring and disincentivizing their diversity, namely how similar or different are a person's experience, competences or traits with regards others \cite{schelenz2021theory}. 

\noindent Secondly, a growing body of literature is focusing on bias and bias management (see, e.g., the extensive work done by \cite{otterbacher2015crowdsourcing, otterbacher2018investigating, orphanou2022mitigating}) in scientific fields such as AI, health, and behavioural studies. 

\noindent Finally, the way in which personal data is treated is not risk-free, especially when considering aspects such as consumer privacy, non-transparent legal regulation, or even bias in the programming. Examples are the processing of data for the purpose of advertising conducted by Google or Facebook \cite{esteve2017business}, but also all the documented cases of "untrustworthy" AI (see e.g., the cases related to face recognition \cite{buolamwini2018gender}). 

Although many strategies to mitigate the risks associated with non-diversity-aware approaches to data have already been proposed, such as explainability \cite{xu2019explainable} or the report on Ethics guidelines for trustworthy AI \cite{eutrustworthy}, we believe that, following \cite{boyd2012critical} suggestion, a broader (and radical) approach should be taken.

This is the reason why we suggest the development of an end-to-end research infrastructure (RI)\footnote{According to \cite{euri}, RIs are "facilities that provide resources and services for the research communities to conduct research and foster innovation in their fields".} that enables trustworthy diversity-aware data within a citizen science community.

\noindent Data management is a complex and multidisciplinary process, ranging from ethical and legal to social sciences and AI. Furthermore, it involves various phases, from collection to preparation up to distribution and reuse (see, e.g., \cite{corti2019managing}). Furthermore, data is used in numerous fields, both for research and innovation. Being RIs pivotal for developing research areas as they consolidate both technologies and methodologies (considering, for example, standards or guidelines), we believe that an end-to-end RI is necessary, to support the researchers or developers within the whole data management process.

\noindent Then, we consider diversity-aware data, in order to represents the uniqueness of people within their context. An analogous term is Big Thick Data, \cite{bornakke2018big}, which aims to combine Big (Thin) Data, which are usually high in volume but provides little or no contextual information \cite{fielding2008sage}, as can be data coming from smartphone sensors (e.g., GPS location, WiFi connection, app usage); and Thick Data, which are contextual data provided by the interaction with the person. Thick interactions can concern a variety of interplay, e.g., of the person with her context or with other people and of the person with the machine collecting the data. From the first interplay derives characteristics such as gender, age or nationality, but also deeper aspects, such as emotions or values connected to specific events; from the latter, it derives feedback on the machine usages. Collected together, this information allows to map people diversity in all its aspects.

\noindent In this sense, diversity-aware data not only allows for the representation of the diversity of people, but it is what enables effective Hybrid Human-Artificial Intelligence approaches, where AI adapts to the human via Thick interactions. In the next section, we will focus on applications that allow the collection of Big Thin and Thick Data via smartphone, which is the pervasive tool par excellence.

\noindent Finally, by trustworthy, we mean a structure that not only complies with the guidelines proposed by the EU commission and the General Data Protection Regulation (GDPR, Regulation (EU) 2016/679 of the European Parliament and of the Council of 27 April) but which is able create a relationship of trust with people who provide their data, e.g., through transparent communication, but also acting as an intermediary in defending their interests.

\noindent In this sense, it is crucial the relationship between those who provide the data and those who analyse it. Therefore, to foster the diversity-aware approach, a community of trust need to be created. We propose to follow the consolidated approach of Citizen Science (CS) \cite{haklay2021citizen}, aiming to involve citizens not only in research but also in a shared data culture.

The remainder of the paper is organized as follow. Section \ref{sec-relatedwork} describes the current status on diversity-aware data generation and sharing. Section \ref{sec-RI} presents a solution outline and challenges for developing a diversity-aware following the exemplary case of \cite{livepeople} within the DataScientia ecosystem. Section \ref{sec-conclusion} closes the paper.

\section{The current status on diversity-aware data generation and sharing} \label{sec-relatedwork}
Although an end-to-end RI addressing diversity-aware data doesn't exist yet, there are several technologies and infrastructures that address parts of it. In particular aspects of (i) data collection; (ii) data management and distribution; (iii) involving people in experiments or within a community.

\paragraph{Data collection} Although increasingly fundamental to CS, “sensing technologies [...] is one area that has not yet been harnessed” \cite{o2016intelligent}, both from a theoretical and technological point of view. Considering diversity-aware data collection, of the several configurable data collection applications, few are able to collect them. Many applications, such as Psychlog \cite{gaggioli2013mobile} and Mobile Sensing Platform \cite{place2017behavioral}, collect data only from user interactions, while others, such as \cite{stack}, collect only sensors data. Two applications are particularly relevant, namely AWARE \cite{ferreira2015aware}, which is a complex system often adopted within the ESM framework\footnote{Experience Sampling Method (ESM) is an intensive longitudinal social and psychological research methodology, i.e. designed for reducing social and cognitive bias in data collection, where participants are asked to report on their thoughts and behaviours.} \cite{csikszentmihalyi2014validity}, but it does not provide data management support, and \cite{cslogger}, whose ease of configuration makes it suitable for a CS community, even if not equipped for collecting all the sensors data.

\paragraph{Data management and distribution} The disciplines that deal with personal data often develops RIs to support researchers in the aspects of data management and sharing. For instance, within social sciences, \cite{ukda} and \cite{gesis} provide support for ethics assessment and data management and, alongside with \cite{osf}, they enable the documentation and distribution of high quality survey data. Data distribution is particularly advanced in the healthcare sector, with leading RIs such as \cite{ontario} or \cite{biobank}, which also have played a key role in the management of the Covid19 pandemic.

However, despite the obvious support provided by such infrastructures, they are not end-to-end. Furthermore, they remain tied to individual research communities, not favouring effective interdisciplinary exchange.

\paragraph{Crowd-sourcing vs. Communities} According to \cite{hosseini2015crowdsourcing}, CS has some aspects in common with crowd-sourcing, especially in involving non-expert people in fulfilling research tasks. Examples are participatory sensing \cite{goldman2009participatory} and Mobile Crowd Sensing \cite{ma2014opportunities} and they are particularly relevant as they rely on the pervasiveness of smart devices to collect data on large panel, even though based on people often coming from Western countries, which is a main issue for considering the diversity of people. 

\noindent However, crowd-sourcing considers the participant only as a contributor to the data collection \cite{shirk2012public}, but rather than an active member of a community, as it does not considers an involvement in the research process nor education and information projects. On the contrary, projects like \cite{zooniverse}, \cite{inaturalist} or \cite{cornelllab} can be considered as actual CS communities, even if their focus is on natural sciences and not on the diversity of people and their behaviour, while \cite{scistarter} and \cite{eucitizen} have a broader focus, even though they are not based on an end-to-end RI.

\section{Towards a diversity-aware RI} \label{sec-RI}
To outline a potential solution, we will present the LivePeople case study and discuss some limitations. Even if not yet fully operational, LivePeople contains a set of proposals of services and technologies that covers the main aspects described in Section \ref{sec-relatedwork} for creating an end-to-end diversity-aware RI embedded in a CS community.

\paragraph{Data collection} One of the LivePeople services is a cutting-edge data collection app called iLog \cite{2020-zeni1}\footnote{\cite{2014-PERCOM,2017-SOCINFO,2017-ICSC,2018-PERCOM2} is a list of publications which describe the use of iLog and of iLog collected data in various studies.}, which allows collecting diversity-aware information through the interaction with the person and from all the smartphone sensors. 

\paragraph{Data management and distribution} The whole data management process is ethics and privacy-aware by design, and it is based on a consolidated methodology considering quality standards from the social science domain \cite{corti2019managing}, and following the \cite{fair}. Regarding this latter principle, the RI also focuses on advanced data integration approaches \cite{giunchiglia2021itelos}, which aims to extend the collected data for interdisciplinary reuse.

\paragraph{The CS community} 
LivePeople will be established on a cross-country panel of people, i.e. it will be based on the diversity of people. To ensure trust in the community, people will remain the owners of their data and have the option to donate or sell it in exchange for research and services of interest to them. Ultimately, RI will be community-based and community-led. Not only will the RI provide services to community members in their context, but the community itself will be self-sufficient to create and run new projects and to support and contribute to existing ones.

\paragraph{Limits} Even if part of the RI has already been applied in different projects (e.g., \cite{smartunitn, wenet}), leading to interdisciplinary publications (e.g., \cite{2017-SOCINFO, assi2023complex}), some key aspects of LivePeople are not yet consolidated or validated. Examples are (i) the usability of the iLog app by non-expert users, such as citizens; (ii) the validation of data management outcomes to effective reuse of resources - both in terms of data quality and their interoperability; (iii) the lack of a panel that can be consulted on demand and incentive strategy that guarantees high collaboration from members within the community.

\section{Conclusion} \label{sec-conclusion}
In this paper we argued how datafication is affecting data management and data quality creating bias in their reuse. Then we showed how the current status of diversity-aware data generation and sharing platform are not suitable for the purpose of creating trust and quality data, and we presented a former solution, considering some of its constrains. 

\section*{Acknowledgments}
The work is funded by the \emph{``WeNet - The Internet of Us"} Project, funded by the European Union (EU) Horizon 2020 programme under GA number 823783.

\bibliographystyle{unsrt}
\bibliography{bibliography}
\end{document}